\documentclass[showpacs,twocolumn,pre,superscriptaddress]{revtex4}

\usepackage{graphicx}

\usepackage{amssymb}
\usepackage{amsmath}
\usepackage{amsfonts}
\usepackage{mathrsfs}

\renewcommand{\epsilon}{\varepsilon}
\newcommand{\figurewidth}{0.45\textwidth}

\begin{document}
\title{Polymer translocation through a nanopore under a pulling force}
\author{Ilkka Huopaniemi}
\affiliation{Laboratory of Physics, Helsinki University of Technology, 
P.O. Box 1100, FIN-02015 HUT, Espoo, Finland}
\author{Kaifu Luo }
\altaffiliation[]{
Author to whom the correspondence should be addressed}
\email{luokaifu@yahoo.com}
\affiliation{Laboratory of Physics, Helsinki University of Technology, 
P.O. Box 1100, FIN-02015 HUT, Espoo, Finland}
\author{Tapio Ala-Nissila}
\affiliation{Laboratory of Physics, Helsinki University of Technology, 
P.O. Box 1100, FIN-02015 HUT, Espoo, Finland}
\affiliation{Department of Physics, Box 1843, Brown University, 
Providence, RI 02912-1843, U.S.A.}
\author{See-Chen Ying}
\affiliation{Department of Physics, Box 1843, Brown University, 
Providence, RI 02912-1843, U.S.A.}

\date{\today}

\begin{abstract}

We investigate polymer translocation through a nanopore under a 
pulling force using Langevin dynamics simulations. We concentrate 
on the influence of the chain length $N$ and the pulling force 
$F$ on the translocation time $\tau$. The distribution of $\tau$ 
is symmetric and narrow for strong $F$. We find that $\tau\sim 
N^{2}$ and translocation velocity $v\sim N^{-1}$ for both moderate 
and strong $F$. For infinitely wide pores, three regimes are observed 
for $\tau$ as a function of $F$. With increasing $F$, $\tau$ is 
independent of $F$ for weak $F$, and then $\tau\sim F^{-2+\nu^{-1}}$ 
for moderate $F$, where $\nu$ is the Flory exponent, which finally 
crosses over to $\tau\sim F^{-1}$ for strong force. For narrow pores, 
even for moderate force $\tau\sim F^{-1}$.
Finally, the waiting time, for monomer $s$ and monomer $s+1$ to exit 
the pore, has a maximum for $s$ close to the end of the chain, in 
contrast to the case where polymer is driven by an external force 
within the pore.

\end{abstract}


\maketitle
\section{Introduction}
The transport of biopolymers through nanopores occurs in many 
biological processes, such as DNA and RNA translocation across 
nuclear pores, protein transport through membrane channels, 
and virus injection~\cite{Alberts,Darnell,Rabin}.
Due to various potential technological applications, such as 
rapid DNA sequencing~\cite{Meller3}, gene therapy and controlled 
drug delivery~\cite{Chang}, recently a number of experimental
~\cite{Kasianowicz,Aktson,Meller2,Henrickson,Meller,Sauer,Li,
Gershow,Chen,Storm}, theoretical~\cite{Storm,Simon,Sung,Park,
diMarzio,Muthukumar,Muthu2,Lubensky,Slonkina,Ambj,Metzler,Ambj2,
Ambj3,Baumg,Chuang,Kantor,Milchev,Luo1,Luo2} and numerical
~\cite{Chuang,Kantor,Milchev,Luo1,Luo2,Huo,Luo3,Chern,Loebl,
Randel,Lansac,Kong,Farkas,Tian,Zandi} studies on polymer translocation 
have been carried out.

In 1996, Kasianowicz \textit{et al}.~\cite{Kasianowicz} demonstrated
that single-stranded DNA and RNA molecules can be driven through the
$\alpha$-hemolysin channel in a lipid bilayer membrane under an 
electric field, and the polymer length can be characterized.
In addition, Li \textit{et al.}~\cite{Li} and Storm \textit{et al.}
~\cite{Storm} showed that a solid-state nanopore could also be used 
for similar experiments.

Inspired by the experiments~\cite{Kasianowicz,Meller,Storm}, a
number of recent theories ~\cite{Storm,Sung,Muthukumar,Baumg,
Ambj3,Chuang,Kantor,Milchev,Luo1,Luo2} have been developed for the
dynamics of polymer translocation. 
Particularly, the scaling of the translocation time $\tau$ 
with the chain length $N$ is an important measure of the 
underlying dynamics. 
Sung and Park~\cite{Sung} and
Muthukumar~\cite{Muthukumar} considered equilibrium entropy of the
polymer as a function of the position of the polymer through the
nanopore. Standard Kramer analysis of diffusion through this entropic 
barrier yields a scaling prediction of $\tau \sim N^2$ for the field 
free translocation. 
For the forced translocation, a linear dependence of 
$\tau$ on $N$ was suggested, which is in agreement with some experimental
results~\cite{Kasianowicz,Meller} for $\alpha$-hemolysin channel.
However, as Chuang \textit{et al}.~\cite{Chuang} noted, 
the quadratic scaling behavior for the field free translocation 
cannot be correct for a self-avoiding polymer. 
The reason is that the translocation time is shorter than the 
equilibration time of a self-avoiding polymer, 
$\tau _{equil} \sim N^{1 + 2\nu }$, where $\nu$ is the
Flory exponent~\cite{de Gennes,Doi}. 
According to scaling theory, they showed that for large $N$, 
translocation time scales approximately in the same manner 
as equilibration time.
For the forced translocation, Kantor and Kardar~\cite{Kantor} 
provided a lower bound for the translocation time that scales as 
$N^{1+\nu}$, by considering the unimpeded
motion of the polymer. 
Most recently, we investigated both free and forced translocation 
using both the two-dimensional fluctuating bond model with 
single-segment Monte Carlo moves~\cite{Luo1,Luo2} and Langevin 
dynamic simulations~\cite{Huo,Luo3}. For the free translocation, 
we numerically verified $\tau \sim N^{1 + 2\nu}$ by considering 
a polymer which is initially placed in the middle of the 
pore~\cite{Luo1,Huo}. 
For the forced translocation, we found a crossover
scaling from $\tau\sim N^{2\nu}$ for relatively short polymers to
$\tau\sim N^{1+\nu}$ for longer chains~\cite{Luo2,Huo}. In addition,
we also found that this crossover scaling remains unaffected
for heteropolymer translocation~\cite{Luo3}. The predicted short 
chain exponent $2\nu\approx1.18$ in 3D agrees reasonably well with 
the solid-state nanopore experiments of Storm \textit{et al.}~\cite{Storm}.

Polymer translocation involves a large entropic barrier, and thus
most polymer translocation phenomena require a driving force, such
as an external electric field used in above mentioned experiments.
However, one can also envisage the use of other forces, such as a
pulling force. With the development of manipulation of single
molecules, polymer motion can be controlled by optical
tweezers~\cite{Gerland,Farkas}. This gives a motivation to study the
translocation in which only the leading monomer experiences a
pulling force. In addition, a new sequencing technique based on a
combination of magnetic and optical tweezers for controlling the DNA
motion has been reported~\cite{Sean}. Therefore, it is of great
importance to theoretically investigate the polymer translocation
under a pulling force. Kantor and Kardar~\cite{Kantor} have
considered the scaling of $\tau$ with $N$ and  with the pulling
force $F$. They have also tested the scaling behavior in a MC
simulation study of the fluctuation bond model. However, as
discussed below, this model is only valid for moderate pulling
forces. For strong pulling forces the scaling of $\tau$ with $F$ is
different and needs to be studied carefully. To this end, in this
paper we investigate polymer translocation through a nanopore under
a pulling force using Langevin dynamics simulations.

\section{Model and method} \label{chap-model}
In our numerical simulations, the polymer chains are modeled as
bead-spring chains. The excluded volume effects and van der Waals
interactions between all pairs of beads are modeled by a repulsive
LJ potential as:
\begin{equation}
U_{LJ} (r)=\left\{  \begin{array}{ll}  4\epsilon \left[ \left(\frac
{\sigma}{r}\right)^{12}-\left(\frac{\sigma}
{r}\right)^6  \right]+\epsilon, & r\le 2^{1/6}\sigma;\\
0, &  r>2^{1/6}\sigma,
\end{array}\right.
\end{equation}
where $\sigma$ is the diameter of a bead, and $\epsilon$ is the
depth of the potential. Nearest neighbor beads on the chain is
connected via a FENE spring with a potential $U_{FENE}
(r)=-\frac{1}{2}kR_0^2\ln(1-r^2/R_0^2)$, where $r$ is the separation
between consecutive beads, $k$ is the spring constant and $R_0$ is
the maximum allowed separation between connected beads.

In the Langevin dynamics method, each bead is subjected
to conservative, frictional, and random forces
${\bf F}_i^C$, ${\bf F}_i^F$, and ${\bf F}_i^R$, respectively,
with~\cite{Allen}
$m{\bf \ddot {r}}_i = {\bf F}_i^C + {\bf F}_i^F + {\bf F}_i^R$,
where $m$ is the monomer's mass.
Hydrodynamic drag is included through the frictional force, which for 
individual monomers is ${\bf F}_i^F=-\xi {\bf v}_i $, where $\xi$ is 
the friction coefficient, and ${\bf v}_i$ is the monomer's velocity.
The Brownian motion of the monomer resulting from the random bombardment 
of solvent molecules is included through ${\bf F}_i^R$ and can be 
calculated using the fluctuation-dissipation theorem.
The conservative force in the Langevin equation consists of several terms
${\bf F}_i^C=-{\bf \nabla}({U}_{LJ}+{U}_{FENE})+{\bf F}_{\textrm{pulling}}$.
The pulling force is expressed as
\begin{equation}
{\bf F}_{\textrm{pulling}}=F\hat{x},
\end{equation}
where $F$ is the pulling force strength exerted on the first monomer
and $\hat{x}$ is a unit vector in the direction perpendicular to the wall.

In the present work, we consider a 2D geometry where the wall in the $y$ 
direction is described as $l$ columns of stationary particles within
distance $\sigma$ from one another and they interact with the beads
by the repulsive part of the Lennard-Jones potential. Wall particle 
positions do not change during the simulations. The pore is introduced 
in the wall by simply removing $w$ beads from the wall.

\section{Scaling Arguments} \label{chap-results}
\subsection{Stretching extension}

Here, we first briefly recall the main results of scaling analysis
pioneered by Pincus~\cite{Pincus}who considered a polymer under 
traction with two forces, $F$ and $-F$, act on its end. The elongation 
$L(F)$ of the chain maybe written as
\begin{equation}
L(F)=R\phi(\frac{R}{\zeta}),
\label{eq31}
\end{equation}
where $\phi$ is a dimensionless scaling function, $R=N^{\nu}\sigma$, 
denotes the size of the unperturbed coil, and $\zeta$ is the 
characteristic length of the problem, $\zeta=k_{B}T/F$.

For weak forces, such that $F<k_{B}T/N^\nu\sigma$, the response is 
linear, i.e., $\phi(x)\sim x$, which leads to~\cite{de Gennes}
\begin{equation}
L(F)\sim N^{2\nu}\sigma \frac{F\sigma}{k_{B}T}.
\label{eq311}
\end{equation}
For moderate forces $k_{B}T/N^\nu\sigma \le F \le k_{B}T/\sigma$, the 
chain breaks up into a one-dimensional string of blobs of size $\zeta$.
Then the elongation $L(F)\sim N$, which leads to $\phi(x)\sim x^{(1-\nu)/\nu}$.
Thus, one obtains~\cite{de Gennes,Pincus}
\begin{equation}
L(F)\sim N\sigma(\frac{F\sigma}{k_{B}T})^{\frac{1}{\nu}-1}.
\label{eq312}
\end{equation}
Finally for strong forces $F>k_BT/\sigma$, the chain is nearly
fully extended with
\begin{equation}
L(F)\sim N\sigma.
\label{eq313}
\end{equation}

In the following, we use similar arguments to analyze the extension
of the tethered chain pulled with a constant force through a viscous
medium. The geometrical impediment due to the finite width of the pore
is neglected here. For clarity, we assume that the pulling force acts
on the last monomer $N$. Without hydrodynamic interactions the force
acting on segment $n$ is given by
\begin{equation}
F_{n} =\xi \sum_{i=1}^{n} v_{i}.
\label{eqFn}
\end{equation}
where $v_{i}$ is the velocity of the $i$th segement. At steady state 
when inertia can be neglected compared with the frictional force, 
we assume $v_{i}=v$ and thus 
\begin{equation}
F_{n}=n\xi v.
\end{equation}
Under this physical picture, we stress that the pulling force $F$ 
equals $F_N =N\xi v$. Here, we encounter a situation in which the 
tension $F_n$ is segment-dependent and where the stretching of the 
chain is not uniform ~\cite{Brochard,Blumen1,Blumen2}.

To this end, we generalize Eqs. (\ref{eq311}) and (\ref{eq312}) to 
this situation, following Brochard-Wyart~\cite{Brochard}, who considered 
the nonuniform deformation of tethered chains in uniform solvent flow.
Let $\zeta_n =k_B T/F_n$ be the $n$-dependent size of the Pincus blobs 
and $x_n$ the position of the $n$th monomer with respect to the last 
monomer in the direction of the pulling force. For weak forces at $n$, 
i.e., $F_n <k_{B}T/N^\nu\sigma$, the local elongation at site $n$ is
\begin{equation}
dx_n\sim n^{2\nu-1}\sigma \frac{F_n \sigma}{k_{B}T}dn,
\label{eq32}
\end{equation}
while for moderate forces, $k_{B}T/N^\nu\sigma \le F_n \le k_{B}T/\sigma$,
\begin{equation}
dx_n\sim \sigma(\frac{F_n\sigma}{k_{B}T})^{\frac{1}{\nu}-1}dn.
\label{eq33}
\end{equation}
Integrating Eqs. (\ref{eq32}) and (\ref{eq33}) over $n$, one finds the 
deformation obeys
\begin{equation}
L(F)\sim N^{2\nu+1}\sigma \frac{\xi v\sigma}{k_BT}
\label{eq411}
\end{equation}
for weak forces and
\begin{equation}
L(F)\sim N^{1/\nu}\sigma(\frac{\xi v\sigma}{k_{B}T})^{\frac{1}{\nu}-1}
\label{eq412}
\end{equation}
for moderate forces.
From the scaling picture, the sizes $\zeta_n$ of the Pincus blobs obey
$\zeta_n \sim 1/F_n \sim 1/n \sim x^{-\nu}$. This can be compared with 
the case of a tethered chain subjected to a uniform solvent 
flow~\cite{Brochard}. The blob size decreases in the pulling force 
direction, resulting in a trumpetlike shape. For strong forces, the 
last part of the chain in fully stretched while its free end still
shows Pincus-type behavior, corresponding to the regime called stem 
and flower ~\cite{Brochard}.

Note that Eqs.(\ref{eq411}) and (\ref{eq412}) can be obtained by simply 
replacing $F$ in Eqs.(\ref{eq311}) and (\ref{eq312}) by an effective 
force $F_{eff} =N\xi v$, which means Eqs.(\ref{eq311}) and (\ref{eq312}) 
are still applicable when one sets
$\zeta=\zeta_{eff} =k_B T/F_{eff}$.

\subsection{Translocation time}
To examine $\tau$ as a function of $N$ under the same constant
pulling force $F$, we need to know $L(F)$ as a function of the
pulling force $F$, not the drag force $\xi v$ on monomer. To use
Eqs. (\ref{eq411}) and (\ref{eq412}) we need to use relation 
$F=N\xi v$, which are the same as Eqs.(\ref{eq311}) and (\ref{eq312})
although the microscopic pictures are different.

Kantor and Kardar~\cite{Kantor} have presented a scaling argument 
for the unimpeded translocation, which serves as a lower bound for 
the true translocation through the nanopore. The polymer travels a 
distance $L(F)$)during the translocation process. The translocation 
velocity scales as $F/N$ since the force is applied to one monomer 
only. Thus the translocation time should depend on $N$ and $F$ as 
$\tau\sim\frac{L(F)}{v(F)}$~\cite{Kantor}. For moderate forces i.e., 
$k_BT/N^\nu\sigma \le F_n \le k_BT/\sigma$, we have from the scaling 
of $L(F)$ in Eq. (\ref{eq412})
\begin{equation}
\tau\sim\frac{L(F)}{v(F)}\sim N^{2}F^{-2+\frac{1}{\nu}}.
\label{eq512}
\end{equation}

\begin{figure}
 \includegraphics*[width=\figurewidth]{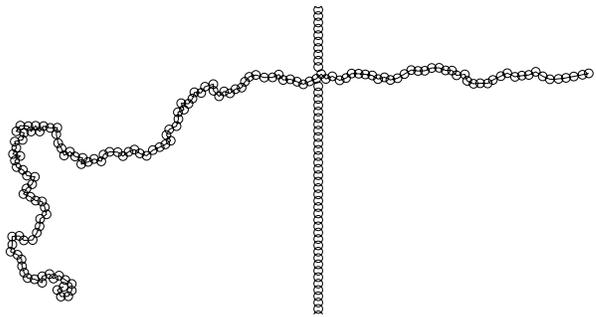}
\caption{A typical configuration of a polymer of length $N=300$ pulled 
by a force of strength $F=5$ during translocation process.}
 \label{Fig1}
\end{figure}

This scaling relation for moderate force is the same as the one
obtained earlier ~\cite{Kantor}. We can now extend this approach to
both weak and strong forces. For weak forces, the translocation time
scales according to Eq.(\ref{eq411}) as
\begin{equation}
\tau\sim\frac{L(F)}{v(F)}\sim N^{1+2\nu}.
\label{eq511}
\end{equation}
This scaling behavior is the same as that for translocation in the
absence of forces~\cite{Chuang,Luo1,Huo}, in disagreement with recent 
claims ~\cite{Panja}.
For strong pulling forces, the polymer becomes fully stretched and
thus obtains a qualitatively different spatial configuration. Such a
configuration is shown in Fig. \ref{Fig1} for a polymer of length
$N=300$ during translocation. In this case, the translocation time
scales as
\begin{equation}
\tau\sim\frac{L(F)}{v(F)}\sim N^{2}F^{-1}.
\label{eq513}
\end{equation}
Eqs. (\ref{eq512}) and (\ref{eq513}) show that $\tau \sim N^2$ for 
both moderate and strong forces. From Eqs.(\ref{eq511}), (\ref{eq512}) 
and (\ref{eq513}), $\tau$ as a function $F$ have three regimes with 
increasing $F$.

\section{Results and Discussions}

In our simulations, the parameters are $\sigma=1$, $k_B T=1.2\epsilon$, 
and the time scale is given by $t_{LJ}=(m\sigma^2/\epsilon)^{1/2}$, which 
is in order of ps. The friction is set as $\xi=0.7m/t_{LJ}$. For the FENE 
potential, we use~\cite{Holm} $R_0=2\sigma$, $k=7\epsilon/\sigma^2$.
Typically, $k_B T/\sigma=4$ pN for a chain with Kuhn length $\sigma=1$ nm 
at room temperature 295 K, and the time scale is about 11.28 ps for monomer 
mass $m=312$ amu. The scale of the pulling force $F$ is $\epsilon/\sigma$, 
which is about 3.3 pN. The Langevin equation is integrated in time by a method 
described by Ermak and Buckholtz~\cite{Ermak} in 2D. For the pore, we set 
$w=3\sigma$ and $l=\sigma$ unless otherwise stated.

To create the initial configuration, the first monomer of the chain
is placed in the entrance of the pore. The polymer is then let to
relax to obtain an equilibrium configuration. In all of our
simulations we did a number of runs with uncorrelated initial
states. The translocation time is defined as the time interval
between the entrance of the first segment into the pore and the exit
of the last segment. The estimate for the translocation time was
obtained by neglecting any failed translocation and then calculating
the average duration of the successful translocations. Typically, we
average our data over 1000 independent runs. In this section, we
only investigate the translocation under the consant pulling force.
For the case with the constant pulling velocity, we will examine it
in detail in future.

\subsection{Translocation time distribution}
\begin{figure}
  \includegraphics*[width=\figurewidth]{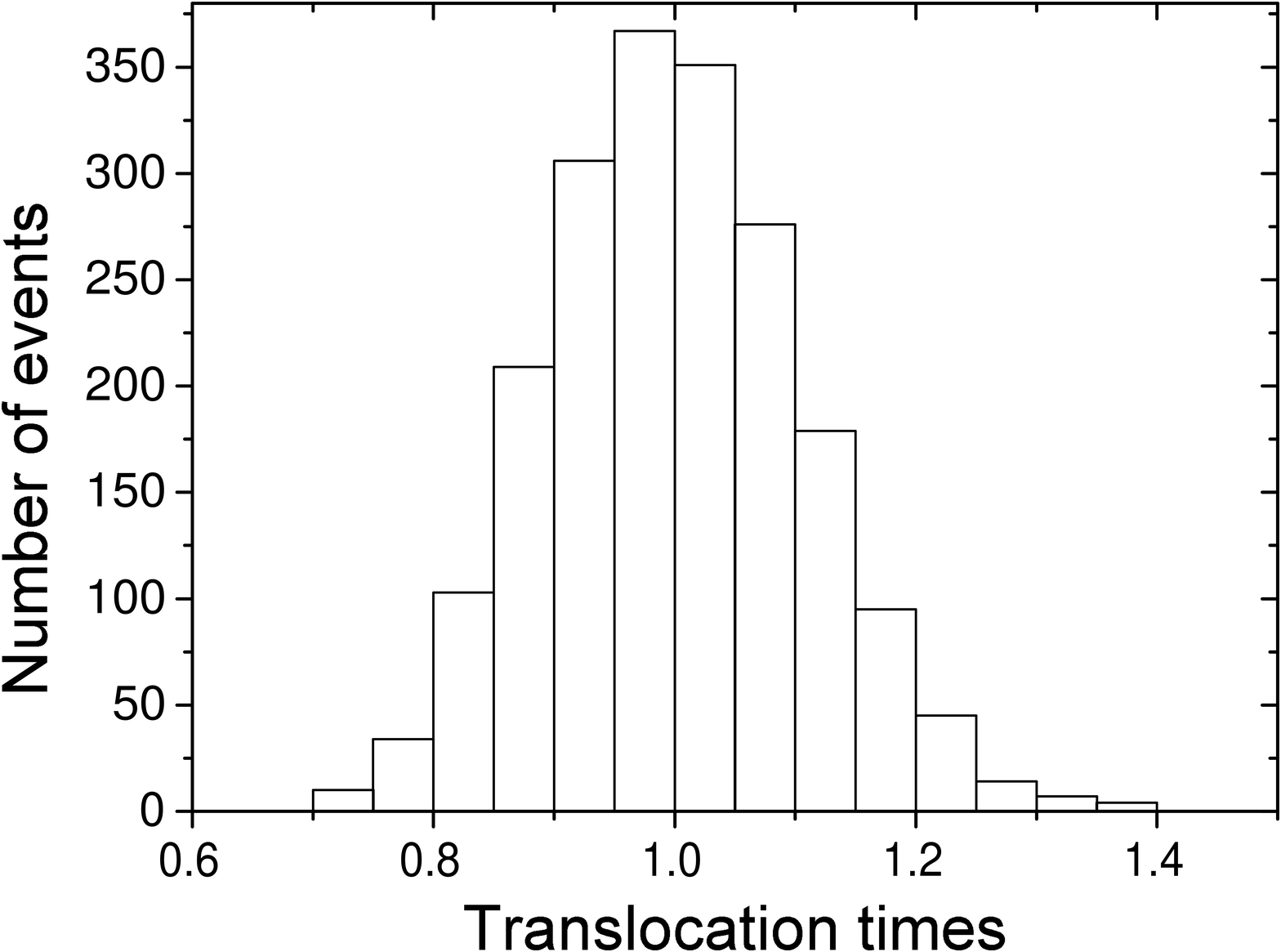}
\caption{The distribution of 2000 translocation times for a chain of 
length $N=100$ under the pulling force of strength $F=5$. Here, the 
translocation times are normalized by their average value. }
 \label{Fig2}
\end{figure}

The distribution of translocation times for a polymer of length 
$N=100$ pulled with a force $F=5$ is presented in Fig. \ref{Fig2}. 
The histogram obeys Gaussian distribution. This distribution has 
a qualitatively different shape compared to that for the free 
translocation case, where the corresponding distribution is 
asymmetric, wider and has a long tail~\cite{Chuang,Huo}. However, 
this distribution is quite similar to that for driven translocation 
under an electric field, in that it is narrow without a long tail 
and symmetric~\cite{Kantor,Huo}. The stronger the pulling force, 
the narrower the distribution becomes. As a consequence of this 
distribution, the average translocation time $\tau$ is well defined 
and scales in the same manner as the most probable translocation 
time. Of course, if a weak enough pulling force is used, we still 
can observe the long tail.

\subsection{Waiting time}
\begin{figure}
  \includegraphics*[width=\figurewidth]{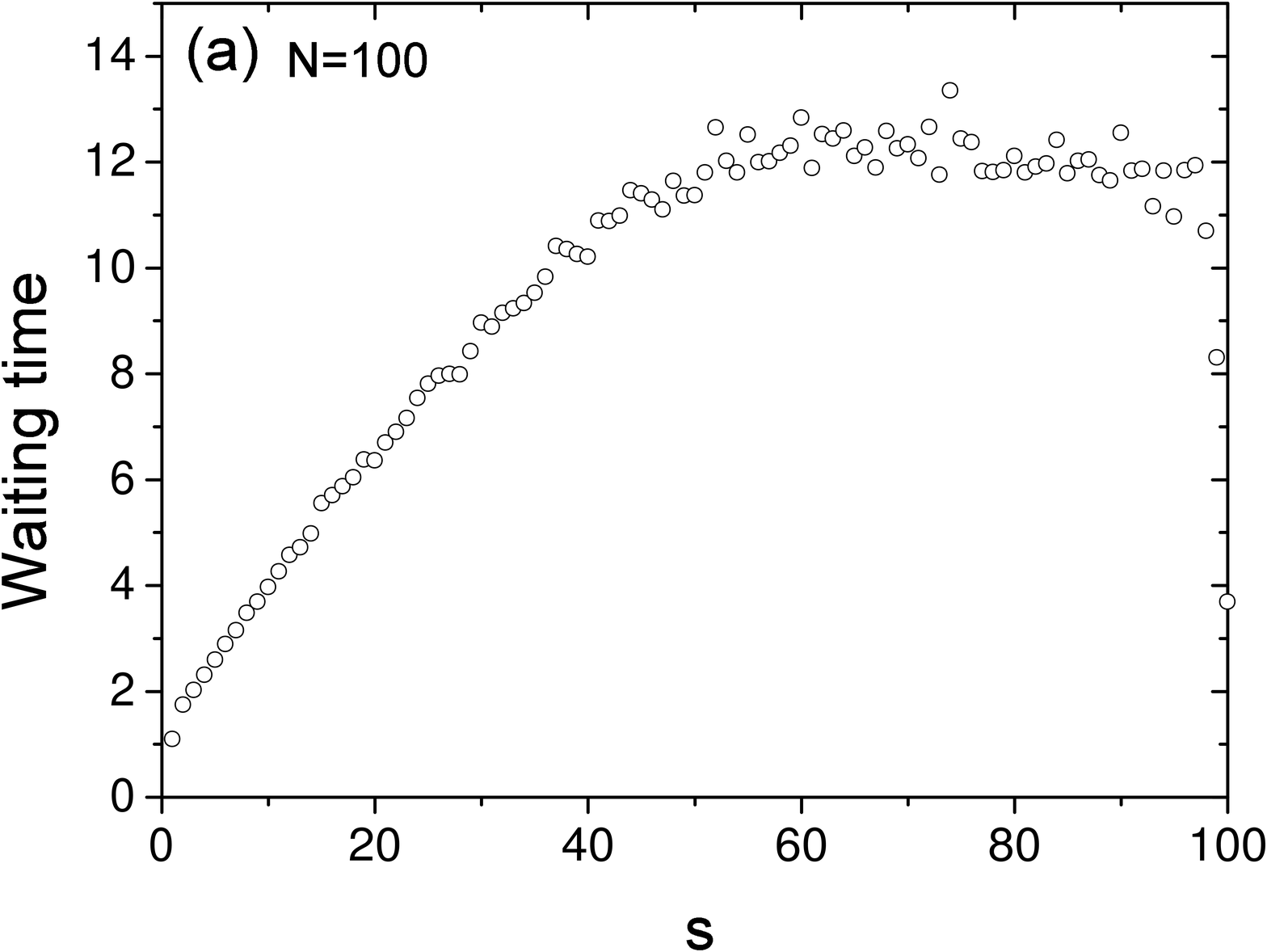}
  \includegraphics*[width=\figurewidth]{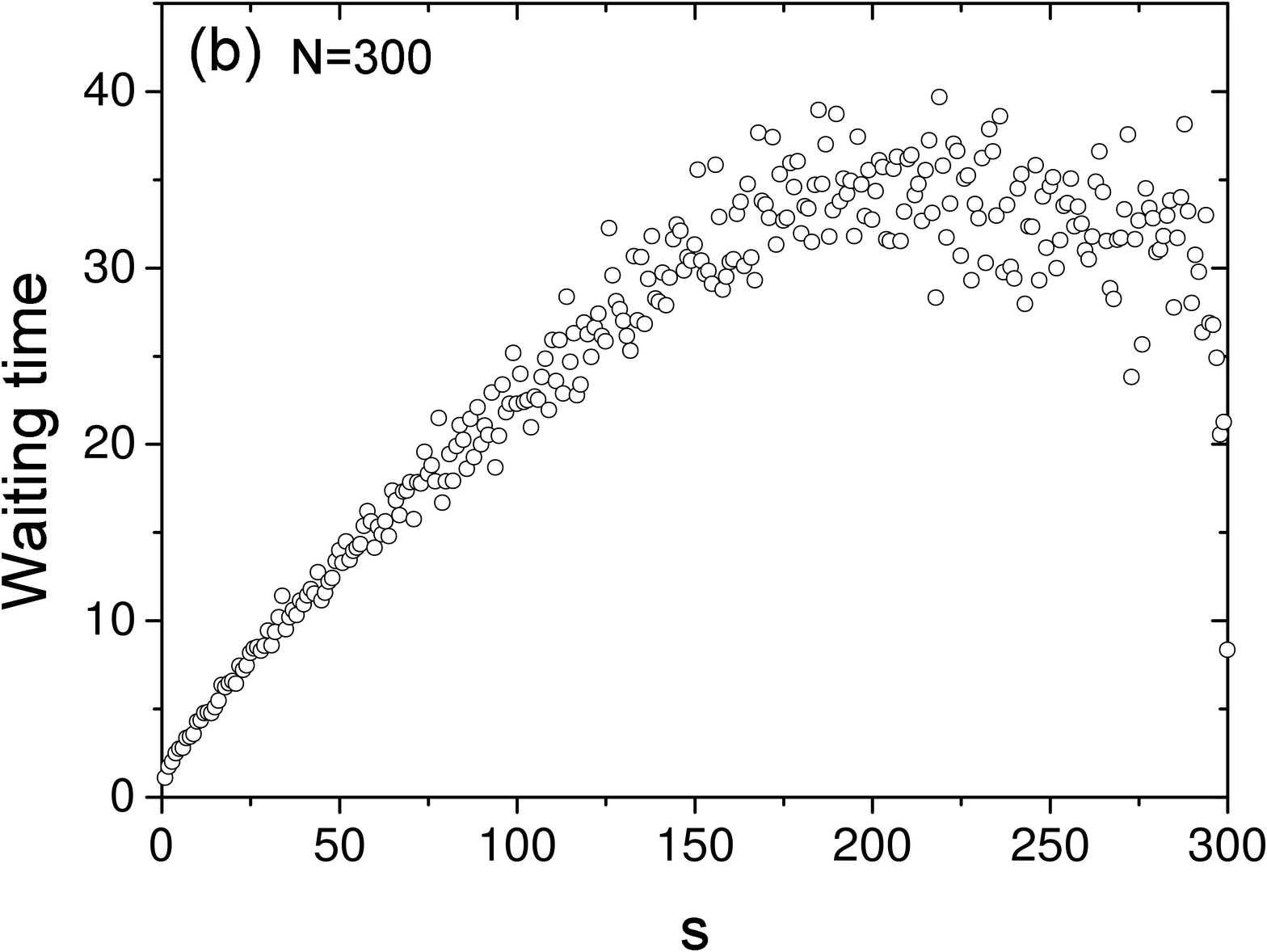}
\caption{Waiting times for a polymer of
(a) length $N=100$ with $F=5$, (b) length $N=300$ with $F=5$.}
 \label{waitingN100}
\end{figure}

The dynamics of a single segment passing through the pore during 
translocation is an important issue. The non-equilibrium nature 
of translocation has a considerable effect on it. We have numerically 
calculated the waiting times for all monomers in a chain of length 
$N$. We define the waiting time of monomer $s$ as the average time 
between the events that monomer $s$ and monomer $s+1$ exit the pore. 
In our previous work~\cite{Luo2,Huo} for translocation under an electric 
field in the pore, we found that the waiting time depends strongly 
on the monomer positions in the chain. For short polymers, such as 
$N=100$, the monomers in the middle of the polymer need the longest 
time to translocate and the distribution is close to symmetric. 
However, for a polymer of length $N=300$, it's approximately the
220$^{\textrm{th}}$ monomer that needs the longest time to translocate 
on the other side of the pore. The waiting times for chains of length 
$N=100$ and $N=300$ under pulling forces are presented in 
Fig. \ref{waitingN100}. As compared to the electric field driven 
case~\cite{Luo2,Huo}, the peaks of the waiting times are shifted towards 
the last monomers of the chain, independent of the force.
This can be understood from the fact that when the chain is being pulled 
through the pore, its free energy increases due to loss of configurational 
entropy. For short chains, this leads to the chain motion to slow down 
almost monotonically until the chain entropies on both sides of the pore 
roughly balance each other. For long chains, the entropy of the pulled 
part eventually exceeds that of the remaining part of the chain and a 
maximum in waiting time appears close to the end of the chain. 

\subsection{Translocation time as a function of the chain length}
\begin{figure}
  \includegraphics*[width=\figurewidth]{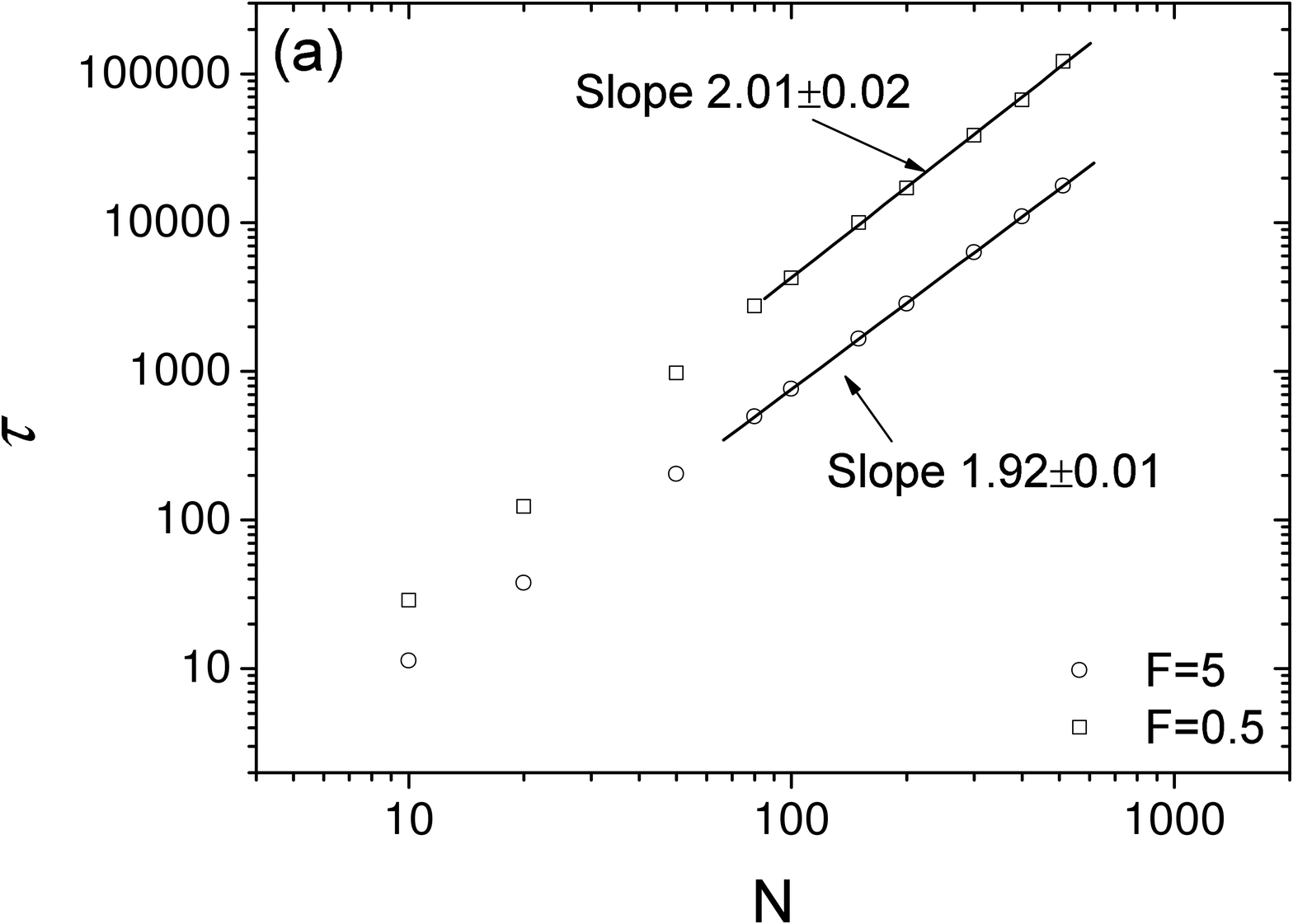}
  \includegraphics*[width=\figurewidth]{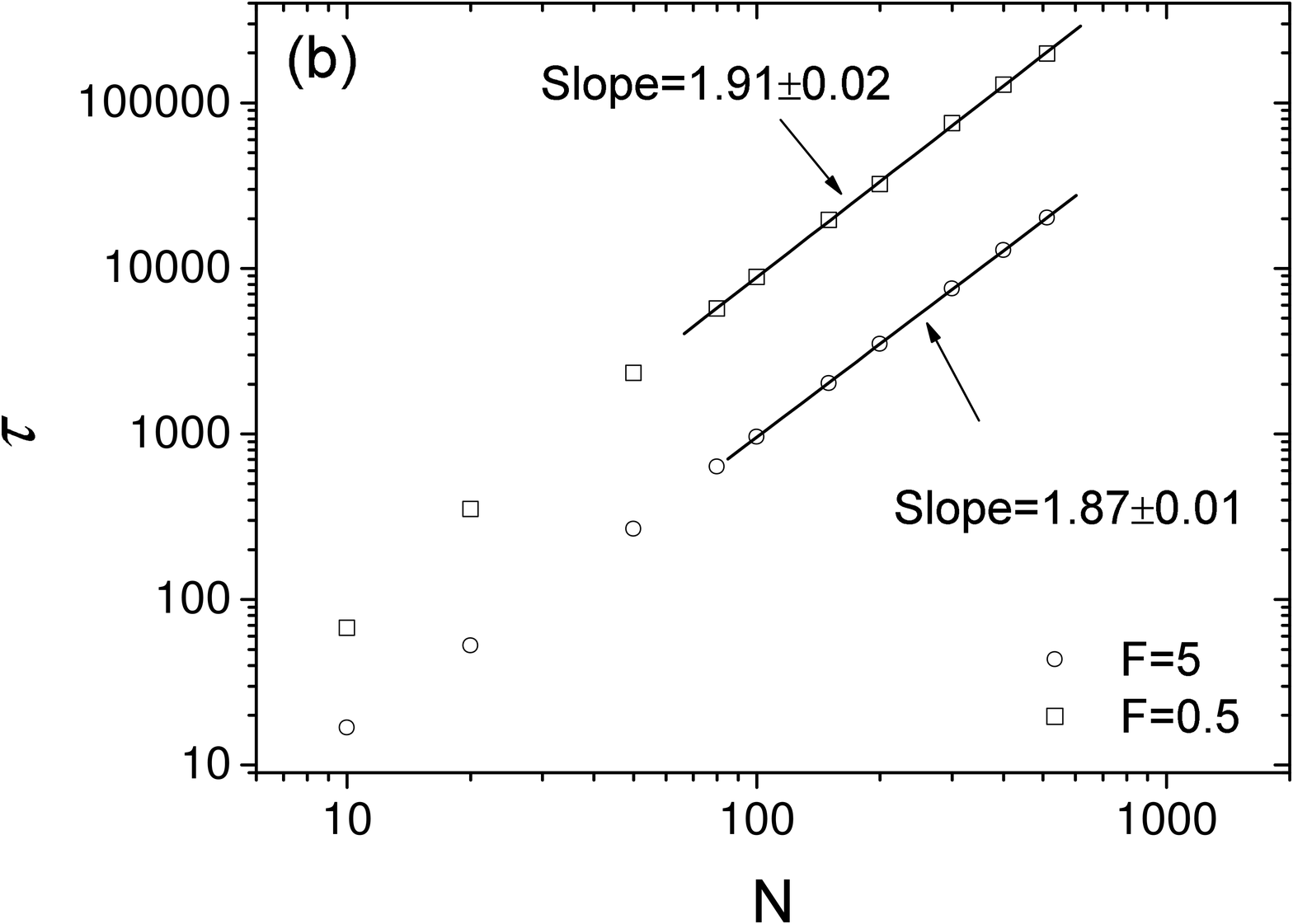}
\caption{The translocation time as a function of polymer length $N$ for
(a) an infinitely wide pore and (b) a pore of finite width. A constant 
pulling force of strength $F=0.5$ and $5$ acts on the first monomer.}
 \label{Fig4}
\end{figure}

As a reference point for comparison, we first check $\tau$ as a
function of $N$ for an infinitely wide pore, as shown in 
Fig. \ref{Fig4}(a). We obtain in this case that $\tau\sim
N^{1.92\pm0.01}$ and $\tau\sim N^{2.01\pm0.02}$ for $F=5$ and $0.5$
respectively. The value for $F=0.5$ is in the moderate force regime
while $F=5$ corresponds to the strong force regime. For both these
regimes, the  theoretical prediction is $\tau\sim N^{2}$, as
 in Eqs. (\ref{eq512}) and (\ref{eq513}). Our numerical results for 
these two pulling forces are in very good agreement with the scaling 
argument predictions. For a pore of finite width, the results are 
shown in Fig. \ref{Fig4}(b).
We get scaling exponents of $1.87\pm0.01$ and $1.91\pm0.02$ for
$F=5$ and $0.5$ respectively. These results are in excellent
agreement with the Monte Carlo simulation results of Kantor and
Kardar \cite{Kantor}, and demonstrate that the scaling arguments for
unimpeded translocation provides a useful estimate for the actual
translocation through the finite size nanopore. 
 
\subsection{Translocation time as a function of the pulling force}
\begin{figure}
  \includegraphics*[width=\figurewidth]{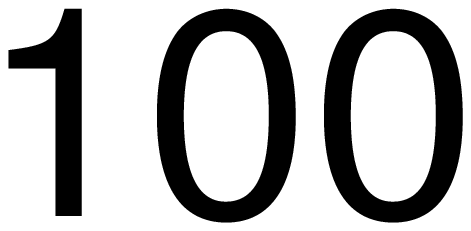}
  \includegraphics*[width=\figurewidth]{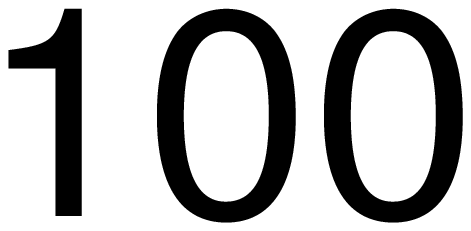}
\caption{Translocation time as a function of pulling force strength for
(a) an infinitely wide pore and (b) a pore of finite width. $\tau$ is an 
average of 1000 runs. Here, $N=100$.}
 \label{Fig5}
\end{figure}

Our theory predicts that there are three regimes in the dependence 
of the translocation time on the pulling force, as shown in 
Eqs. (\ref{eq511}), (\ref{eq512}) and (\ref{eq513}). To study
this, we again consider first the unimpeded translocation through an
infinitely wide pore. The numerical results in Fig. \ref{Fig5}(a)
confirm the existence of the three regimes. The translocation time is
independent of $F$ for weak pulling forces, i.e., $F \le 0.3$, which
is indicated in Eq. (\ref{eq511}). With increasing pulling force,
the translocation time scales with the force with an exponent of
-0.67 for $0.3 \le F \le 2$. This result is in good agreement with
the theoretical prediction in Eq. (\ref{eq512}), where $\tau \sim
F^{-2+\frac{1}{\nu}}\sim F^{-0.67}$ in 2D. However, for $2 \le F \le 10$, 
the exponent is -0.84. This shows that 
$L(F)\sim(\frac{Fa}{k_{B}T})^{\frac{1}{\nu}-1}$ is no longer valid
since we are in the strong force regime of $F>k_{B}T/\sigma$.
Instead, we expect $L(F)$ to be almost independent of the force and
correspondingly the translocation time should scale as Eq. (\ref{eq513})
in the limit of a strong force.

For a pore of finite width, the translocation time as a function of
pulling force is presented in Fig. \ref{Fig5}(b). 
For $0.25 \le F \le10 $ we find $\tau\sim F^{-0.94\pm0.01}$. 
These results show that under the restriction of the wall for a pore
of finite width, it is much easier for the chain to become fully
stretched and hence the strong force limit scaling behavior holds
$\tau\sim F^{-1}$ through the entire range of forces applied.

Finally, it is important to note that both in the case of external
field (voltage applied across the pore)~\cite{Luo1,Luo2} and pulling 
force driving the translocation process, there exists a fundamental 
difference between the Monte Carlo results for the lattice fluctuating 
bond model and the continum model considered here in the strong driving 
force limit. In the Monte Carlo study, the microscopic transition rate 
saturates very quickly when the external driving force increases, 
leading to a saturation of the velocity and the translocation 
time~\cite{Luo1,Kantor}. This aspect of the fluctuating bond model 
is unrealistic and does not correspond to the true dynamics of the 
system. The continum model does not suffer from this artifact. 
As seen in Fig. \ref{Fig5}, in the present model the translocation 
time $\tau$ scales as $\tau\sim F^{-1}$ up to the maximum force value 
studied and shows no sign of saturation. Our previous studies of the 
field driven translocation with both Monte Carlo and Langevin dynamics 
show that while the scaling behavior agree in most regimes, the same
difference occurs in the strong force limit.

\section{Conclusions} \label{chap-conclusions}

In this work, we have investigated the dynamics of polymer translocation
through a nanopore under a pulling force using 2D Langevin dynamics 
simulations. We have focused on the influence of the length of the chain 
$N$ and the pulling force $F$ on the translocation time $\tau$.
The distribution of $\tau$ is symmetric and narrow for strong $F$.
We find that $\tau\sim N^{2}$ and translocation velocity $v\sim N^{-1}$ 
for both moderate and strong $F$. For infinitely wide pores, three regimes 
are observed for $\tau$ as a function of $F$. With increasing $F$, $\tau$ 
is independent of $F$ for weak $F$, and then $\tau\sim F^{-2+\nu^{-1}}$ 
for moderate $F$, where $\nu$ is the Flory exponent, which finally crosses 
over to $\tau\sim F^{-1}$ for strong force. For narrow pores, even for 
moderate force $\tau\sim F^{-1}$. Finally, the waiting time, for monomer $s$
and monomer $s+1$ to exit the pore, has a maximum for $s$ close to the end 
of the chain, in contrast to the case where polymer is driven by an external 
force within the pore.

\begin{acknowledgments}
This work has been supported in part by The Academy of Finland
through its Center of Excellence (COMP) and TransPoly Consortium grants.
\end{acknowledgments}

\end{document}